\pdfoutput=1 %arXiv admin added this line
\documentclass{article}%
\usepackage{amssymb}%
\usepackage{amsmath}%
\usepackage{amsfonts}%
\usepackage{graphicx}
\providecommand{\U}[1]{\protect\rule{.1in}{.1in}}
\begin{document}

\title{\textbf{Comment on "Pressure dependence of wall relaxation in polarized
He}$^{3}$ \textbf{gaseous cells." }by W. Zheng, H. Gao, Q. Ye, and Y.Zhang}
\author{A. Petukhov$^{\left(  1\right)  }$, J. Chastagnier$^{\left(  1\right)  }$,
C.M. Swank$^{\left(  2\right)  }$ and R. Golub$^{(2)}$\\$^{\left(  1\right)  }$Institute Laue-Langevin, 38042, Grenoble Cedex, France \\$^{\left(  2\right)  }$Physics Dept, North Carolina State University, Raleigh,
NC, 27695, USA}
\maketitle

\begin{abstract}
The authors have demonstrated a strong linear pressure dependence of the
longitudinal relaxation time for $He^{3}$ at room and cryogenic temperatures
in a given experimental setup. They offer a theoretical explanation of the
effect based on diffusion theory in the bulk and an unusual boundary
condition. We question the physical basis of the boundary condition and
suggest some alternate explanations of the observations.

\end{abstract}

In \cite{gao} the authors have reported on some measurements of the spin
relaxation time, $T_{1}$ in polarized $\ He^{3}$ in various cells at room and
cryogenic temperatures. All the results show a strong, linear pressure
dependence of $T_{1},$ which the authors interpret as a general property of
wall relaxation.

The authors recognized that their observation is inconsistent with other
experimental data for low-pressure cells where no apparent pressure dependence
was observed. They mentioned that these cells usually have hundreds or even
thousands of hours of T1 at 1atm, which is considerably longer than the cells
used in \cite{gao} (with $T_{1}$ less than 20 hours) so that there are
probably different relaxation mechanisms operating.

The authors present an explanation of their result based on diffusion theory
(their equ. 3), with the boundary condition given by their equation (5).
According to diffusion theory the boundary condition is
\begin{equation}
D\left.  \frac{\partial\rho\left(  r,t\right)  }{\partial r}\right\vert
_{r=R}=-\frac{v}{4}\beta\rho\left(  r,t\right) \label{1}%
\end{equation}
where $D$ is the diffusion constant, $\beta$ the (dimensionless) loss
probability (in this case a depolarization probability) per bounce and $v$ is
the average velocity of the diffusing particles. (See, e.g equation (10) of
\cite{grebenk}). Zheng \emph{et al }\cite{gao} replace this by their equation
(5) and make the unusual assumption \cite{gaoetal} that $\alpha$ in that
equation is independent of $D$. They do not give any explanation of the
physical meaning of their coefficient alpha.

It is this which leads to their predicted pressure dependence. For a
discussion of the boundary condition (\ref{1}) in relation to the diffusion
limit of various kinds of random walks see \cite{erban}. Then (continuing with
the diffusion theory analysis) to first order in $\beta$, the eigenvalue
solutions to their equation (7) is $x_{k}^{2}=3\beta vR/4D+k\pi$,
($k=$integer) implying $1/T_{1}=x_{0}^{2}D/R^{2}=3\beta v/4R\sim\beta/\tau
_{b},$ independent of pressure (at least for a uniform surface, see below).
($\tau_{b}=R/v$ is the time for a ballistically moving particle to traverse
the cell). This pressure independence persists until $\alpha R\sim1 $ in which
case $x_{1}^{2}\gtrsim1.$ Then, we would have $T_{1}\sim\tau_{D}%
\lessapprox1\sec$ \ compared with $\sim$20h for their cell filled with 1bar
3he. (Here $\tau_{D}=R^{2}/D$ is the time for a particle to diffuse across the
cell.) If the observed $T_{1}$ is much larger than $\tau_{D},$ then the theory
predicts it will be independent of pressure (for a uniform surface).

The physical reasons for the usual pressure independence of wall relaxation
are well known to the community, and are explained in the discussion around
equation (1) in Zheng \emph{et al}. The point is that the number of wall
collisions per particle per second for the entire ensemble is independent of
pressure and thus so is the ensemble average wall depolarization rate. This
only breaks down when $\alpha R\sim\beta\tau_{D}/\tau_{b}\sim1$. This means
that the absorption (depolarization) is so strong that it depletes the
polarization density near the walls. In this case the apparent depolarisation
cannot be faster than the diffusion time. This suggests a possible explanation
of the observations of Zheng \emph{et al}. Localized regions of the surface
with a high depolarization probability would deplete the local density
resulting in a linear pressure dependence, while, since they cover only a
fraction of the surface, the resulting $T_{1},$ would not be too short. In the
experiment the cells used at room temperature were made of Rb coated Pyrex and
were connected to an o-ring valve through a 1.5 mm i.d. Pyrex capillary
tubing. In the low temperature cells the o-ring valve was connected to the
cells by a 3mm by \ 68 cm long tube with an 0.8mm id, 3 mm long restriction.

These attachments may be acting as a localized region with very high
depolarization rate while the rest of the cell has a negligible effect on the
apparent relaxation rate.

Other explanations are possible. As pointed out by Zheng \emph{et al,} their
reference [8] ascribes the observed pressure dependence to a pressure
dependence of $\beta$, associated with an adsorbed phase on the surface. While
gradient relaxation at the high pressures used in their experiment also give
$T_{1}\propto p$, \cite{catesschaeffer}, Zheng \emph{et al }$\ $have ruled out
static gradients as a possible cause of the observed behavior. However there
is the possibility that inhomogeneous a-c fields which result in a similar
pressure dependence of $T_{1},$ \cite{cateswhite} may be playing a role. A
further complication is that we have good evidence, in another experiment,
that the Nitrogen, used as a quenching gas in the SEOP, is getting into our
cell at low temperatures (%
%TCIMACRO{\TEXTsymbol{<}}%
%BeginExpansion
$<$%
%EndExpansion
1K) and exerting a large influence on $T_{1}.$ In the authors' reference [10]
the dependence of $T_{1}$ on $He^{4}$ buffer gas pressure amounts to an
increase of 20\% for a total pressure increase from 1200 to 3000 torr, far
from the claim of Zheng \emph{et al }of strong linear dependence. However
these results were also influenced by gradient relaxation to some extent.
Finally, the authors' model would predict a very short relaxation time $\sim
$1s for a pressure $\sim$1mBar which is in strong contradiction with T1$\sim$8
hours observed in the ILL optical pumping station \cite{khand} which has been
operating for more than 10 years. We are currently engaged in an experiment
with $He^{3}$ to repeat the room temperature results. We are using a Pyrex
cell of the same size(6 cm diameter) and the same order of magnitude
$T_{1}(\sim7h)$ at 1 Bar as the cell used by Zheng \emph{et al.}

To date our preliminary results shown in fig.1 show no strong linear pressure
dependence of $T_{1}$ over a range of pressures from 0.15 to 4.0 bar
(comparable to fig. 2 of the authors' work) for a clean Pyrex cell. We will
shortly be publishing this data along with measurements for a Rb coated cell.
A possible weak inverse dependence $(T_{1}\sim1/p)$ may indeed be attributed
to ferromagnetic relaxation \cite{gao}.%

%TCIMACRO{\FRAME{ftbpFU}{3.3856in}{2.6052in}{0pt}{\Qcb{Measured longitudinal
%relaxation rate for He3 in a clean Pyrex cell at room temperature as a
%function of He3 pressure in the cell.}}{}{petdata.jpg}%
%width 3.3856in;  height 2.6052in;  depth 0pt;  original-width 3.2403in;
%original-height 2.4865in;  cropleft "0";  croptop "1";  cropright "1";
%cropbottom "0";  filename 'PetData.jpg';file-properties "XNPEU";}}}%
%BeginExpansion
\begin{figure}
[ptb]
\begin{center}
\includegraphics[
natheight=2.486500in,
natwidth=3.240300in,
height=2.6052in,
width=3.3856in
]%
{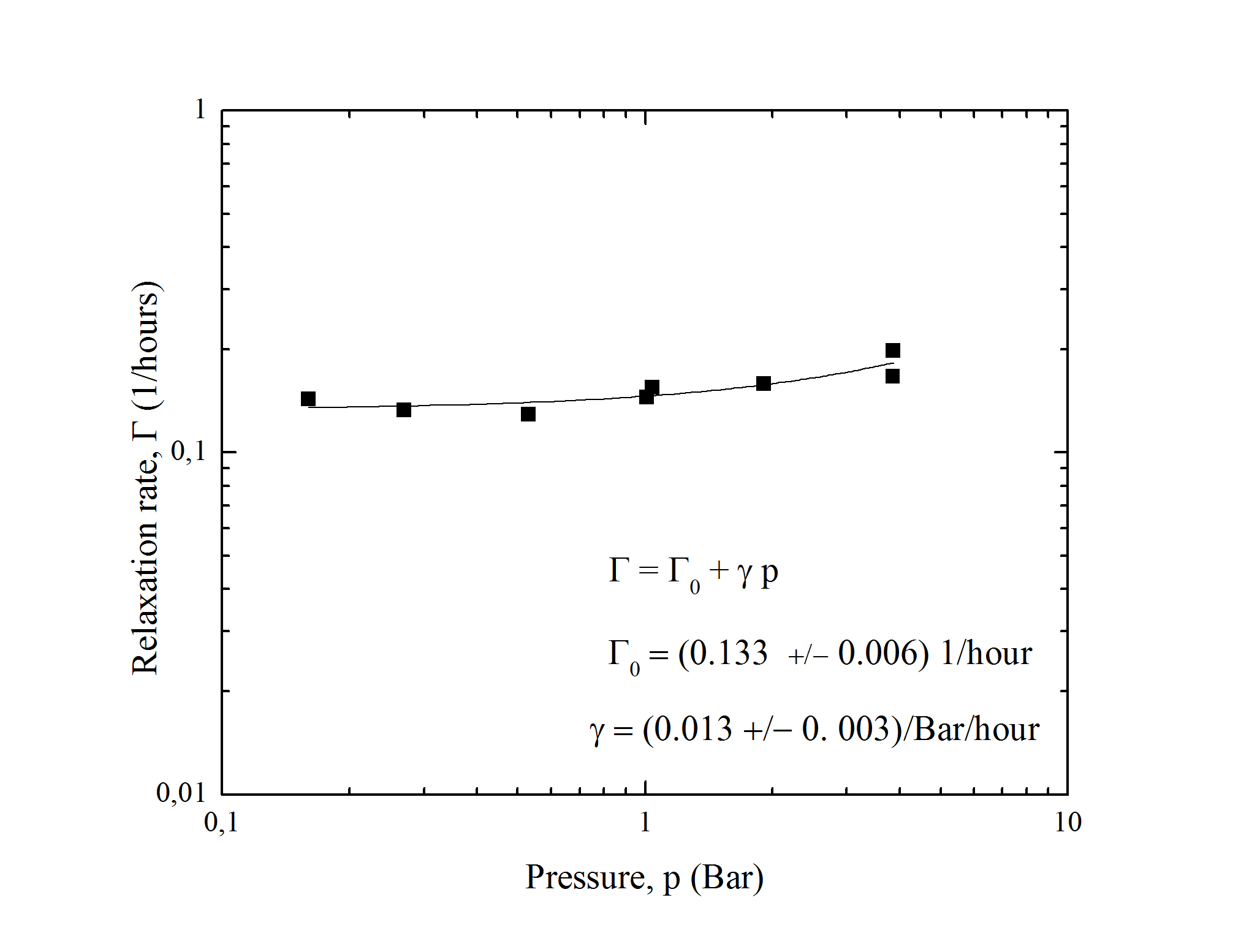}%
\caption{Measured longitudinal relaxation rate for He3 in a clean Pyrex cell
at room temperature as a function of He3 pressure in the cell.}%
\end{center}
\end{figure}
%EndExpansion

Hence, it seems that the observations of Zheng \emph{et al} do not represent a
general property of wall induced relaxation but are the result of \ the
properties of their individual cells and/or other details of their experiment.

Whatever the complexities of the experimental situation, the authors' claim
\cite{gaoetal} that $\alpha$ in their equation (5) is independent of $D$ has
no physical basis. As the equation is written $\alpha$ has the units of
inverse length. The only physical length in the bulk being the mean free path
of the diffusing particles we are back to the diffusion constant,
$D\sim\lambda v$ ($\lambda$ being the mean free path for collisions.) There is
no justification for setting the current proportional to the gradient other
than diffusion theory. The application of the diffusion theory in their
equations (3) and (6) and the denial of its applicability in equation (5) seem inconsistent.

\bigskip
\end{document}